\documentclass[prl,aps,amsfonts,nofootinbib,twocolumn]{revtex4-1}
\usepackage{mathrsfs}
\usepackage{mathtools}
\usepackage{soul}
\usepackage{graphicx}
\usepackage{multirow}
\DeclareMathOperator{\tr}{tr}
\usepackage{color}

\begin{document}

\title{Network complexity and topological phase transitions}


\author{Felipe Torres$^{1,2}$, Jos\'e Rogan$^{1,2}$, Miguel Kiwi$^{1,2}$,
and Juan Alejandro Valdivia$^{1,2}$}

\affiliation{ $^{1}$ Departamento de F{\'\i}sica, Facultad de
  Ciencias,
  Universidad de Chile, Casilla 653, Santiago, Chile 7800024\\
  $^{2}$ Centro para el Desarrollo de la Nanociencia y la
  Nanotecnolog\'\i a, CEDENNA, Avda. Ecuador 3493, Santiago, Chile
  9170124 }

  
\begin{abstract} 

 A new type of collective excitations, due exclusively
  to the topology of a complex random network that can be
  characterized by a fractal dimension $D_F$, is investigated. We
  show analytically that these excitations generate phase transitions
  due to the non-periodic topology of the $D_F>1$ complex
  network.  An Ising system, with long range interactions over such a
  network, is studied in detail to support the claim. The analytic
  treatment is possible because the evaluation of the partition
  function can be decomposed into closed factor loops, in spite
    of the architectural complexity.  This way we compute the
  magnetization distribution, magnetization loops, and the two point
  correlation function; and relate them to the network topology. In
  summary, the removal of the infrared divergences leads to an
  unconventional phase transition, where spin correlations are robust
  against thermal fluctuations.  
    \end{abstract}


\date{\today}  
\maketitle

Phase transitions have attracted significant attention over many
 years. In this context the analytic achievement known as the
Onsager~\cite{Onsager44} solution of the two-dimensional (2D) Ising
model and the Mermin-Wagner theorem (MWT)~\cite{Mermin66,Mermin68}
have been important landmarks in the development of the field.

Here we investigate a complex network with Ising nearest-neighbor
({\it nn}) plus random long-range interactions, and propose a
mechanism that provides a local enhancement of the
correlation length. We find that, as a consequence of the inclusion of 
frustrated long-range interactions, our non-periodic system 
breaks inversion symmetry, due to its fractal nature. Consequently,
 when the effective fractal dimension $D_F > 1$ acquires non-integer
 values, incoherent fluctuations are strongly suppressed.

 During more than half a century the dimensionality restrictions
 imposed by the MWT~\cite{Mermin66,Mermin68,Coleman73,Hohenberg67},
 which for integer $D<3$ is responsible for the suppression of
 correlations by thermal fluctuations, has posed a challenging problem
 in areas like condensed matter and high energy physics. During the
 last two decades the strong dependence of the collective behavior of
 embedded topologies has given impulse to the study of magnetic
 systems on more intricate architectures. This way complex networks
 have become a fertile ground in the study these non-ideal
 systems~\cite{Dorogovtsev08,Reka02,Dorogovtsev02}, since in spite of
 their compact structure, the spatial fluctuations in complex networks
 do give rise to a wide range of critical phenomena, that for $D<3$
 disappear as the fractal dimension of the network approaches an
 integer value.

There is a considerable number of publications on the subject of phase
transitions on complex networks. For example, the ferromagnetic Ising
model on Watts-Strogatz networks displays a ferromagnetic transition
at finite temperature~ \cite{Barrat00, Gitterman00}. This same
phenomenon was also obtained by Monte Carlo simulations, for D= 2 and
3 on {\it small-world} networks, by Herrero~\cite{Herrero02} and by Hong et
al.~\cite{Hong02}. These results are in agreement with mean field
theory. However, numerical simulations by Aleksiejuk et
al.~\cite{Aleksiejuk02}, and predictions based on microscopic
theory by Dorogovtsev et al.~\cite{Dorogovtsev02} and by Leone et
al.~\cite{Leone02}, confirm that the phenomenology associated with
complex networks goes well beyond mean field theory.  Moreover,
Yi~\cite{Yi08, Yi10} studied fluctuations of the Ising
model on a scale free network including an external field orthogonal
to the easy axis~\cite{Seung11}. The critical exponent of this
critical point has been investigated for several
models, such as the Ising-XY in a transverse field~\cite{Botet82,
  Botet83}, and an infinite Ising chain~\cite{Nagaj08, Krzakala08}.

We note that in these systems the underlying commonality is the
emergence of phase transitions produced by the topological structure
of the complex network. Hence, in this manuscript we show, mainly
using analytic tools, that due to the topology of a ({\it small-world})
complex network, the local correlation length can increase due to 
randomly frustrated interactions. Indeed, when the fractal dimension
$D_F >1$ the complexity of the network can yield ``long range''
correlations of the collective excitations.

We demonstrate the effect by means of two approaches: i)~we use a
qualitative description that suggests how the topological complexity
of the network can produce such a phase transition.
This approach, in essence, generalizes Landau's phenomenological
proposal for critical phenomena as they occur in complex
networks; ii)~we construct an Ising model with the spins at
the nodes of the complex network, and they interact through {\it nn}
and long range interactions. The network in this specific case is of a
{\it small-world} type, which has implications that we
will discuss below, for the correlation length and the nature of the
phase transition itself~\cite{Watts-Strogatz}. The partition function
of the model is calculated analytically, so that all the thermodynamic
properties can be evaluated. Below we show, that this phase transition
of the Ising model on such a network is robust
against thermal fluctuations, and consequently that an unconventional
second order phase transition does occur.  Our theory also reproduces
qualitatively the results of Chen et al.~\cite{Chen92} that confirmed,
through simulations of a 2D eight-state Potts model, that the presence
of randomly distributed ferromagnetic bonds changes the phase
 transition from first to second order. The same analysis can be
applied to results that were obtained by Theodorakis et
al.~\cite{Theodorakis12} with a 2D Blume-Capel model embedded in a
triangular lattice, and by Fytas et al.~\cite{Fytas09} for an
antiferromagnetic Ising model with next nearest neighbor interactions. 

The concept of ``long range'' correlations to induce phase transitions
requires some discussion. In general, we expect to observe a phase
transition, and its associated critical phenomena, when the
correlation length becomes large, e.g., effectively of the size of the
system. In the case of the Ising model over a complex network of the
{\it small-world} type, it is also true that the correlation length is about
the size of the system. However, in this case the complex topology
reduces the effective distance between nodes, allowing the system to
become of the size of the correlation length. This is why we have used
a {\it small-world} type network, which allows a large system to have small
average correlation length. This observation should be quite general,
and be applicable to a number of other systems, which we expect on the
basis of the phenomenological approach we present below.

We start our analysis with a qualitative characterization of the
general phase transition problem in complex networks, particularly of
the {\it small-world} type. The phase a system adopts, in thermal
equilibrium, is characterized by spontaneous symmetry
breaking~\cite{Landau} and by the ratio of the order parameter and the
characteristic length of the system. Our main assumption is that the
addition of the {\it small-world} structure (long range interactions) does
not change significantly the correlation length but that, due to
cluster decomposition, it reduces its size-scale and consequently
increases the local correlation. Following Landau~\cite{Landau} we
assume invariance under time reversal inversion and discrete
symmetries to write, neglecting higher order terms of the order
parameter $\psi$, the free energy $\Phi_0({\bf x},T)$ as $\Phi_0({\bf
  x},T)=A(T) |\nabla \psi|^2+B(T) |\psi|^2$, where ${\bf x}$ is the
spatial coordinate, $\psi= \psi_0e^{-x/\xi}$ , $\xi=\sqrt{A(T)/B(T)}$
is the correlation length, and $A$ and $B$ are functions of the
temperature $T$. For $T>T_c$, $A(T) \ll B(T)$, and consequently $\xi
\ll 1$. On the contrary, below the critical temperature $T<T_c$,
$A(T)\gg B(T)$, and $\xi \gg 1$. This way the phase transition is
characterized analytically.  In order to estimate the critical
dimensionality, we use the energy equipartition theorem which,
including spatial fluctuations of the classical ground state, takes
the form
\begin{equation}
\int d^{D}x \,\Phi_0({\bf x}, T<T_c)\geq k_BT \ ,
\label{eq:Equipartition}
\end{equation}
where $D$ is the dimensionality of the system and $k_B$ is
the Boltzmann constant, and the inequality is due to quantum
fluctuations. 

\begin{figure}[h!]
\centering
\includegraphics[scale=0.35]{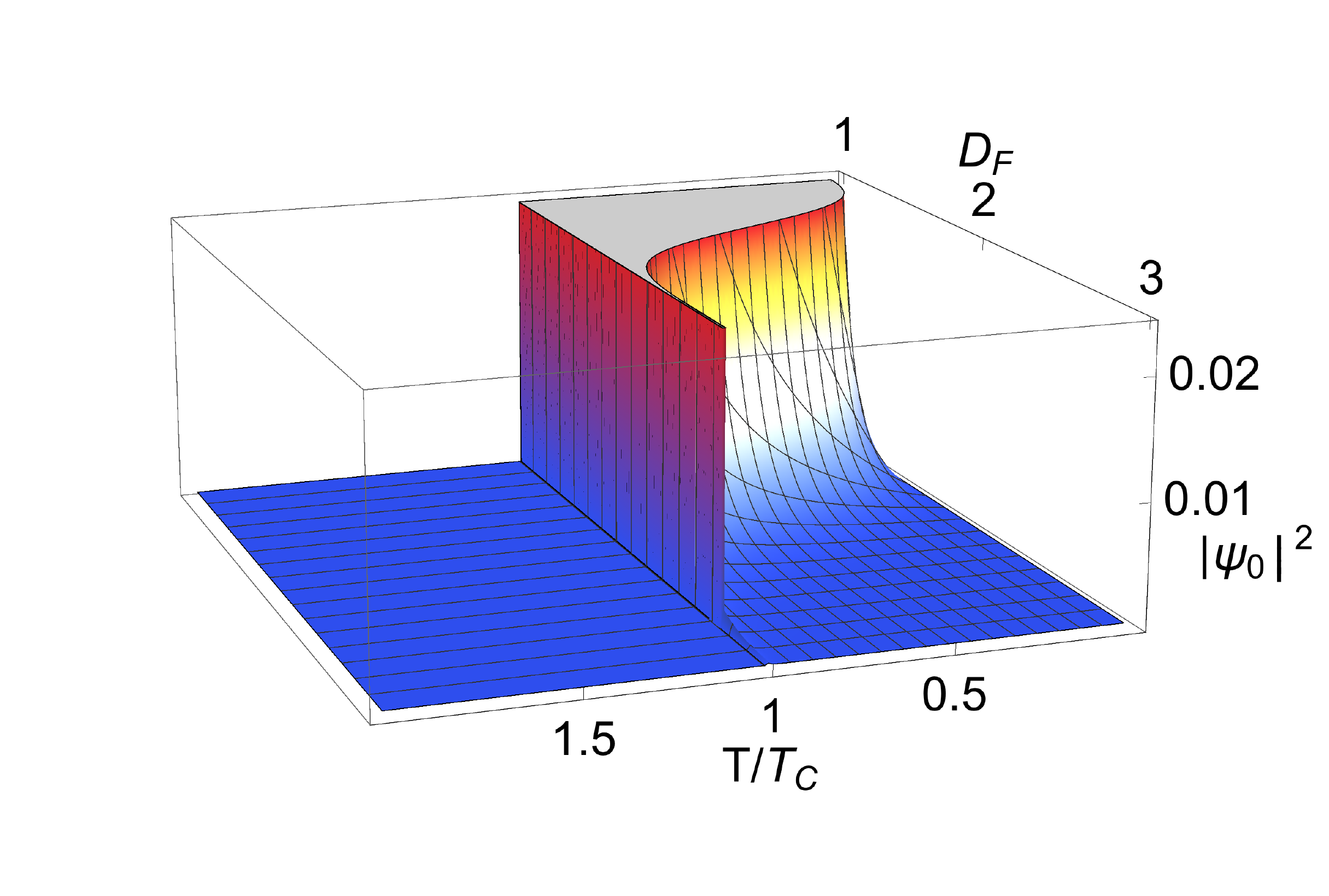}
\caption{(color online). Regularization of infrared divergences when the fractal
  dimension $D_F>1$.}
\label{fig:Order_Parameter}
\end{figure}

Using the explicit expression for the order parameter in
Eq.~(\ref{eq:Equipartition}), we obtain (for $0<T<T_c$)
\begin{equation}
|\psi_0|^2\geq \dfrac{\rho k_BT}{A(T)(\pi
  \xi)^{D}\,\Gamma(D)}\left.\dfrac{k^{D-2}}{D-2}\right|^{k_0}_{k\to0}
\ , 
\label{eq:Divergence}
\end{equation}
where $1/\rho=\int d^{D}k/(2\pi)^D$, $\Gamma(y)$ is the Gamma
function, and $1/k_0$ is the range of the interaction in units of the
{\it nn} distance. This leads to a $\left.1/k\right|_{k\to0}$ infrared
divergence for $D=1$, and a $\left. \ln k\right|_{k\to0}$ divergence
for $D=2$, a result that is equivalent to the MWT. It
is worth mentioning that our phenomenological model recovers results
derived from a microscopic theory, based on the Bogoliubov
inequality~\cite{Torres14}.

Now we focus our attention on how this phenomenological theory can be
applied to a complex network with both {\it nn} plus long range random
links. This system can be characterized in many ways, for example by
an effective fractal dimension or an average path length $L$.  The
effective fractal dimension $D_F$ can be defined as the scaling
$\left<N\right>\sim d^{D_{F}}$ of the average number $\left<N\right>$
of nodes within a radius $d$ of a given node. The average path length
$L$ is the average distance $L=\frac1{2N(N-1)}\sum_{i\neq j}d_{ij}$,
where $d_{ij}$ is the shortest path length between node $i$ and node
$j$, and $N$ is the number of nodes. Therefore, by means of a
box-counting procedure, the inversion symmetry breaking may be
characterized by an effective fractal dimension, which is expected to
couple the free energy to a current density axial vector term, defined
as
\begin{equation}
{\bf J}({\bf x})=i(\psi({\bf
  x})\nabla\psi^*({\bf x})-\psi^*({\bf x})\nabla\psi({\bf x}))/2\ .
\end{equation}
This axial vector ${\bf J}({\bf x})$ is the simplest term that one can
incorporate to break inversion symmetry. The above equation implies
that $|\nabla \psi|^2\sim J^2/|\psi|^2$, where $J({\bf x}) = {|\bf
  J}({\bf x})|$. This is a higher order term that, for $T < T_C$, can
be neglected compared to $\Phi_1({\bf x}, T)=\gamma(T)J({\bf x})$,
where $\gamma(T)\sim k_B(T_C-T)$.  When this last term is included in
the free energy, we obtain
\begin{equation}
|\psi_0|^2\geq \dfrac{\rho k_BT}{\gamma(T)(\pi
  \xi)^{D}\,\Gamma(D)}\left.\dfrac{k^{D-1}}{D-1}\right|^{k_0}_{k\to0}
\,.
\label{eq:DivergenceRandom}
\end{equation}
This way the infrared divergences are removed for $D>1$ since
$|\psi_0|^2\sim k^{D-1}/(D-1)$, so that the restrictions on phase
transitions for $D<3$ are removed. With the insight gained from
dimensional regularization of the phenomenological approach, we now
show how random exchange interactions lead to long range magnetic
order. Our starting point, or unperturbed Hamiltonian, is a 1D
periodic system ({\it i.e.}  a ring with only {\it nn} interactions),
and the random link interactions are added perturbatively {\it i.e.}
$J_0/J\ll 1$, the expansion in momentum eigenstates is valid. But, this
random link distribution brings about a dimensionality change, from
integer to fractal $(D\to D_{F})$, and the breaking of the periodicity
of the system (which is essential for the MWT to hold).  In
Fig.~(\ref{fig:Order_Parameter}) we show the temperature and fractal
dimension $D_F$ dependence of the order parameter, for $k_0\sim 1$ and
assuming that the elementary excitation density, of correlation length
$\xi^{D_F}/\rho$, remains constant. It is thus apparent that the
introduction of random links removes the infrared divergences, and
that phase transitions are in principle now allowed.

We now illustrate the procedure with a
ring Ising model, in order to analyze and illustrate the
characteristics of this phase transition.  As is well known, in the
absence of random links, the 1D Ising model obeys the MWT and does not
exhibit phase transitions~\cite{Mermin66} due to the periodicity of
the interactions. To the 1D ring Ising model composed of $N$ nodes,
with only {\it nn} interactions, we add the possibility of long range
interactions with other particles on the ring, thus generating a complex
network. The Ising Hamiltonian with long range random interactions is
given by
\begin{equation}
\mathscr{H}=-\sum_i\bigg(J\sigma_i\sigma_{i+1}+h\sigma_i
+ J_0\sigma_i\sigma_{r_i}\bigg),
\label{eq:Ising}
\end{equation}
where $\sigma_i=\pm1$; $J>0$ and $J_0$ are the exchange constants
between {\it nn} and long range neighbors, respectively, and $|J_0| <
|J|$; $g$ is the gyromagnetic ratio; $h=g\mu_BH_0$; $\mu_B$ is the
Bohr magneton; and $H_0$ is the uniform applied magnetic field.
For simplicity, we consider a single particle, namely $r_i$,
  connected to particle $i$. Consequently, the first term in
Eq.~(\ref{eq:Ising}) describes the {\it nn} exchange, the second the
Zeeman interaction, and the third the exchange between particle $i$
and the one located at $ r_i$. The effective fractal dimension, and
the average path length, grow linearly with $\ln N$, consistent with a
{\it small-world} topology, which increases the connectivity of
the network locally and enhances the collective behavior of the
system.


\begin{figure}[h!]
\centering
\includegraphics[scale=0.34]{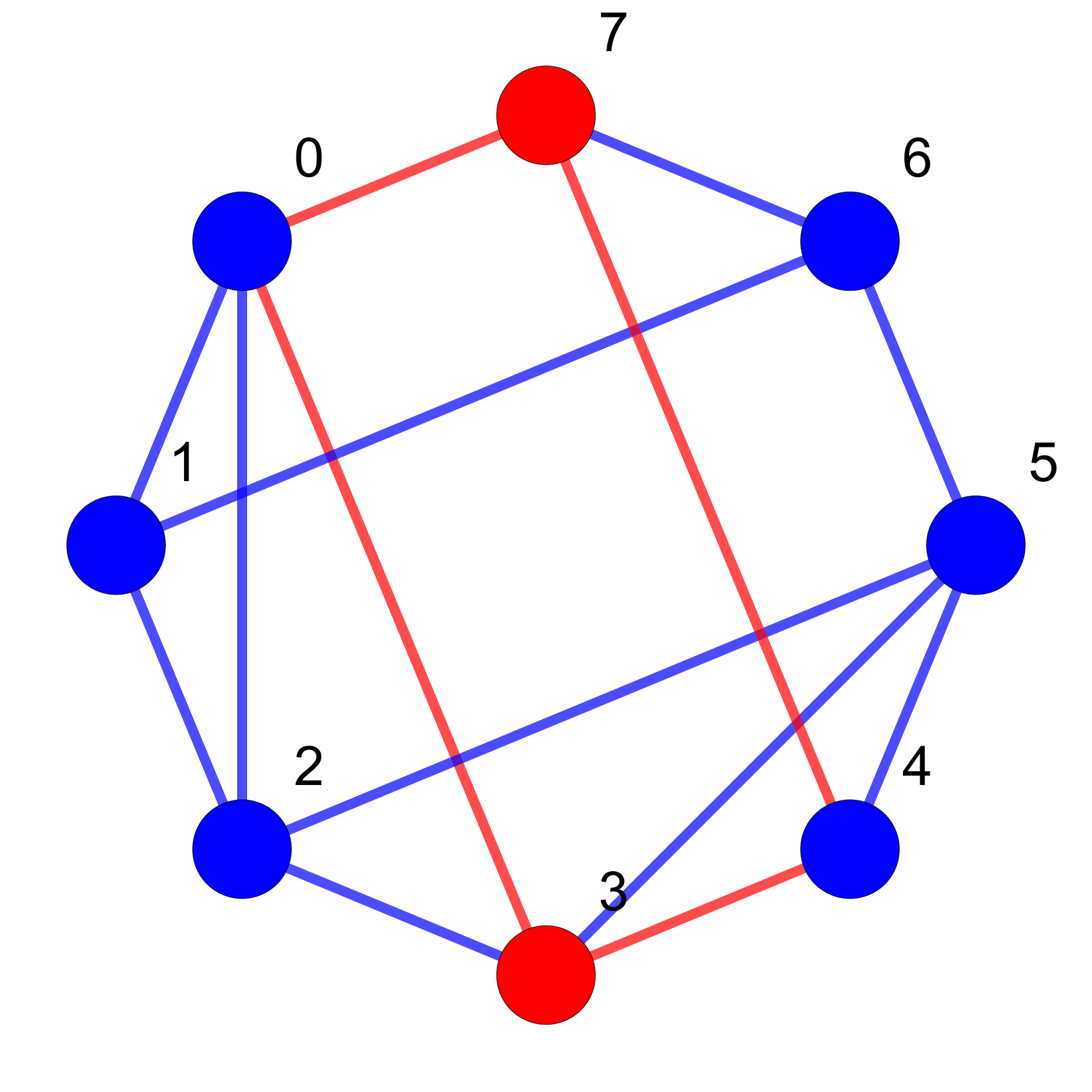}
\caption{(color online). Geometric representation of the decomposition
  into closed factor loops. The loop factors are nodes 064251 (blue)
  and 37 (red).  Following Eqs.~(\ref{eq:matrix}) and
  (\ref{eq:Partition_Function}) they correspond to the factorization
  $M= [M_{06}M_{64}M_{42}M_{25}M_{51}M_{10}] [M_{37}M_{73}] = \tr(M^6)
  \tr(M^2)$ }
\label{fig:Transfer}
\end{figure}

In order to calculate effects  due to the random links on the phase
transition we implement the transfer matrix method, to calculate
  analytically the partition function. We start defining the
matrices
\begin{eqnarray}
T_{\sigma_i,
  \sigma_j}&=&\left<\sigma_i\right|e^{\epsilon_{i,j}/k_BT}\left|\sigma_j\right>,
\nonumber\\ 
R_{\sigma_i, \sigma_j}&=&
\left<\sigma_i\right|e^{\Delta_{i,j}/k_BT}\left|\sigma_j\right>,\nonumber\\ 
M_{\sigma_i, \sigma_j}&=&\sum_{\sigma_\ell}T_{\sigma_i, \sigma_{\ell}}R_{\sigma_{\ell}, 
\sigma_j},
\label{eq:matrix}
\end{eqnarray}
where $J_0\neq 0$, $\epsilon_{i,j} =
J\sigma_i\sigma_j+h(\sigma_i+\sigma_j)/2$ and $\Delta_{i, j} =
J_0\sigma_i\sigma_j$. $J$ and $J_0$ describe the
{\it nn} and long range interactions, respectively.  $\Delta_{i, j}$
describes the interaction between the magnetic moment at sites $i+1$
and $r_i$ through site $i$. These definitions allow to write the
partition function of the system as
\begin{eqnarray}
Z&=&\sum_{\{\sigma_1, \sigma_2,...\}}\sum_{\{\sigma'_1,
  \sigma'_2,...\}} \prod_i T_{\sigma_{i+1},\sigma'_i}R_{\sigma'_i 
, \sigma_{r_i}}\nonumber\\
&=&\sum_{\{\sigma_1, \sigma_2,...\}} \prod_i M_{\sigma_{i+1}
, \sigma_{r_i}}
=\prod_{k}\tr(M^{n_k}) 
\label{eq:Partition_Function}
\end{eqnarray}
where $n_k$ is the number of vertices of each closed loop of nodes
that is formed as we follow the random links along the network.
Eq.~(\ref{eq:Partition_Function}) specifies the cluster decomposition
of the transfer matrix in closed loop factors. This decomposition is
illustrated in Fig.~(\ref{fig:Transfer}) for an $N=8$ configuration.

The $J_0=0$ limit is recovered recalling that $Z=\tr
(T^{N})=\lambda_+^N+\lambda_-^N$, where $\lambda_+>\lambda_-$
are the $T$ matrix eigenvalues. In the $N\to\infty$ limit $Z =
\lambda_+^N$. However, when long range interactions are
incorporated the spatial fluctuations are decomposed
into closed loop factors, which in the $N\to\infty$ limit 
regularize the infrared divergences because we now have a
multiplication of traces which correspond to the closed loop
  factors, so that $\tr(M^{n_0})=\lambda_+^{n_0}+\lambda_-^{n_0}$
must be kept in full, for some $n_0={\min \{ n_k\}}\ll N$.

\begin{figure}[h!]
\begin{center}
\includegraphics[scale=0.43]{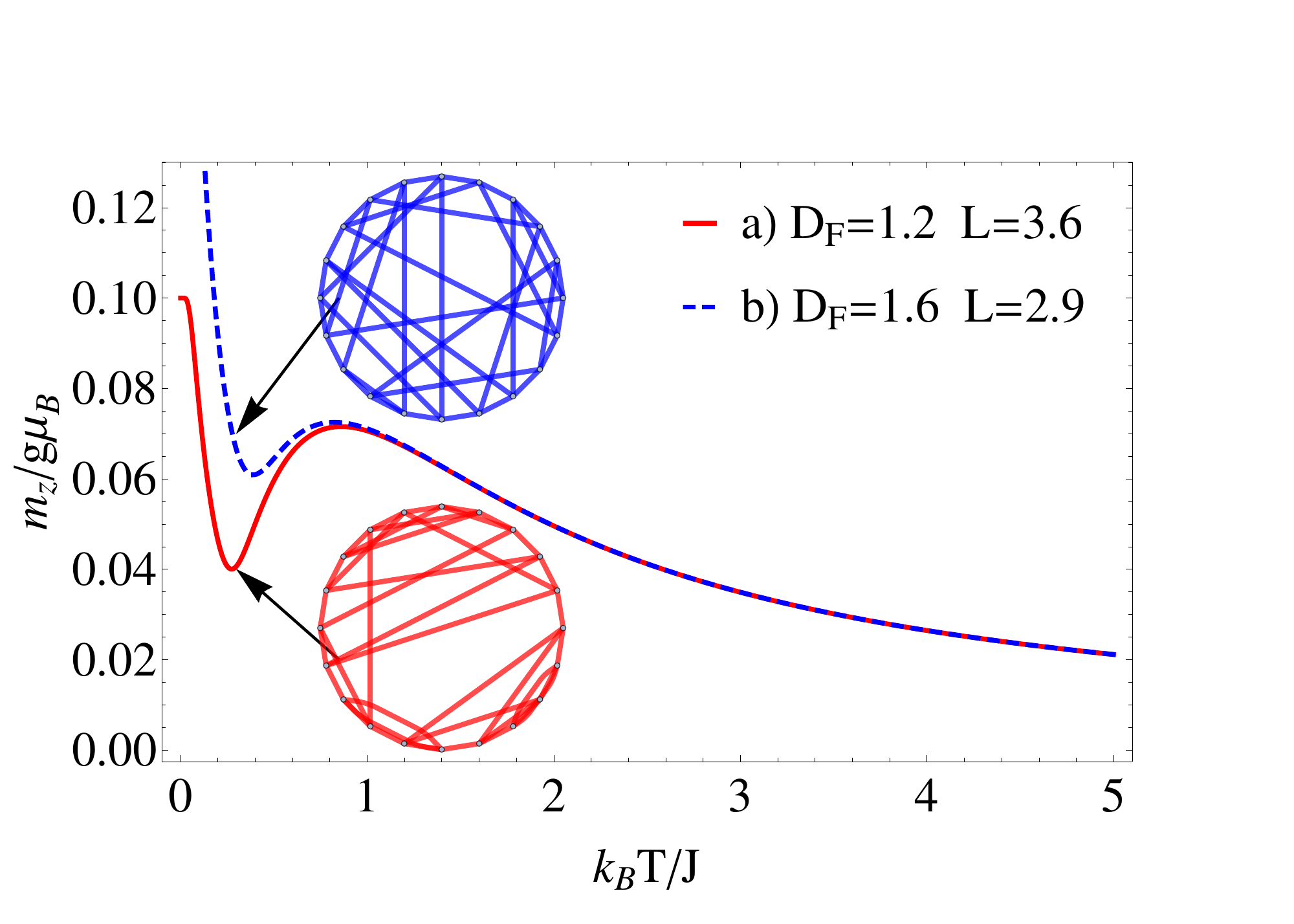}
\caption{Magnetization per particle $m_z$ as a function of temperature
  for two small $D_F$ values, and for two different average path
  lengths $L$.  }
\label{fig:MvsT}
\end{center}
\end{figure}

\begin{figure}[h!]
\includegraphics[scale=0.3]{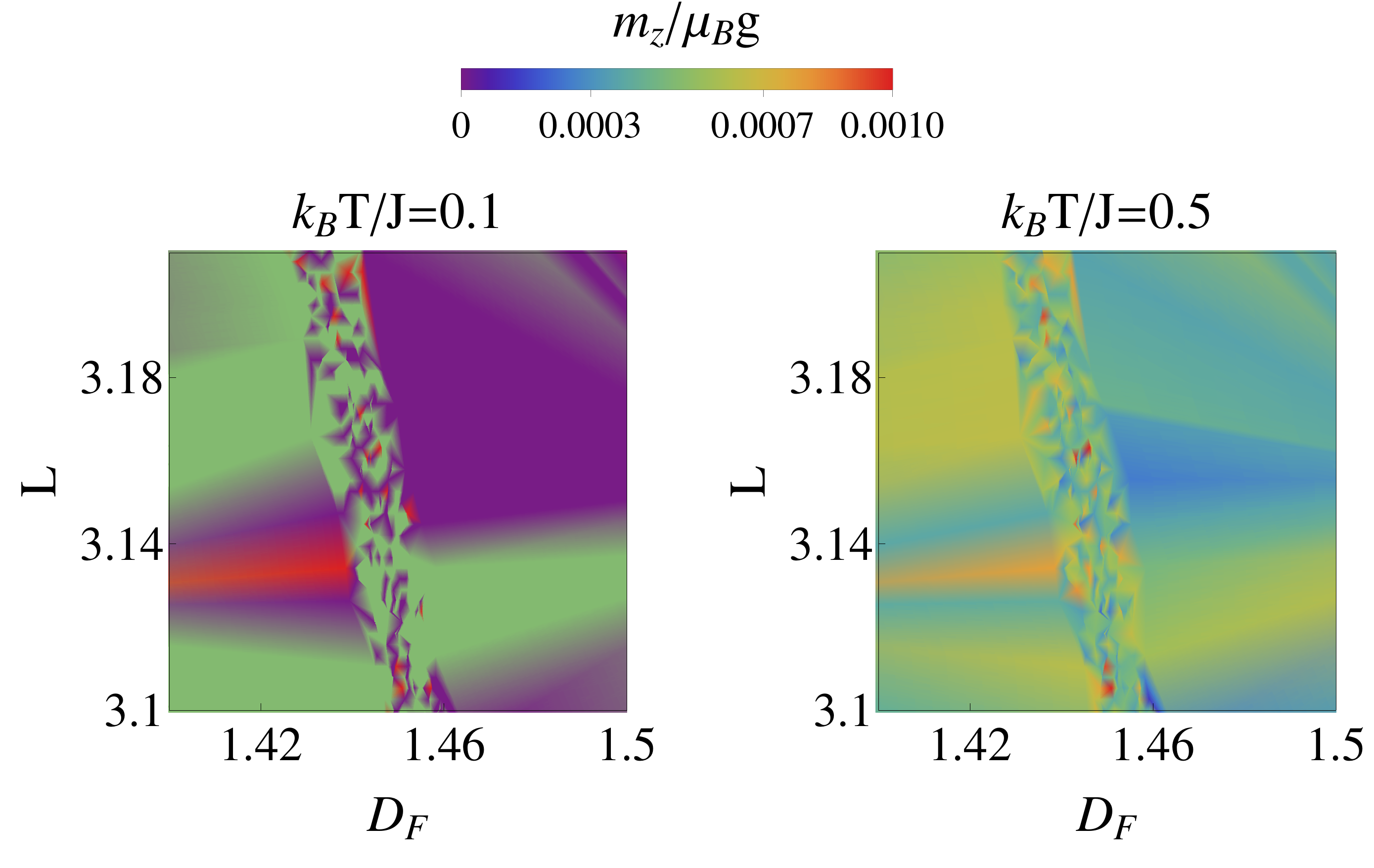}
\caption{Magnetization per particle $m_z$ averaged over
  $10^3$~{\it small-world} network configurations as a function of
  temperature $k_BT/J$, for $h/J=10^{-3}$. The color-scale indicates
  the magnetization magnitude in $g\mu_B$ units.  Each one of this
  {\it small-world} $N=20$ node networks corresponds to a 1D Ising
  ring with one long range link per node.}
\label{fig:PhaseDiag}
\end{figure}

In the absence of an applied field ($h=0$) and at finite temperatures,
the average magnetic moment per particle $m_z=(N\beta)^{-1}\partial
\ln Z/\partial h$ vanishes for the Ising model, which implies that
long range order is destroyed by thermal fluctuations. However, as the
degree of interconnectivity between nodes grows, the fluctuations
become suppressed. This in turn leads to an unconventional continuous
phase transition for $J_0/J<0$, {\it i.e.} when the magnetic
configuration is determined by a competition of short and long range
interactions of different sign. Due to the local frustration this
symmetry breaking could be expected to generate a transition to a spin
glass arrangement. However, as a consequence of the {\it small-world}
network topology, below the critical temperature of a locally ordered
magnetic structure distribution is favored. This is illustrated
  in Fig.~(\ref{fig:MvsT}), where we display the magnetization {\it
    vs.} temperature for two $D_F$ values, and two different average
  path lengths. We remark that the dependence of $m_z$ on $T$ has a
  highly nonlinear dependence on the fractal dimension $D_F$. This
  fact is also reflected in the $m_z\ne 0$ values shown in
  Fig.~(\ref{fig:PhaseDiag}), in the applied magnetic field $h
  \rightarrow 0$ limit.

\begin{figure}
\begin{center}
\includegraphics[scale=0.43]{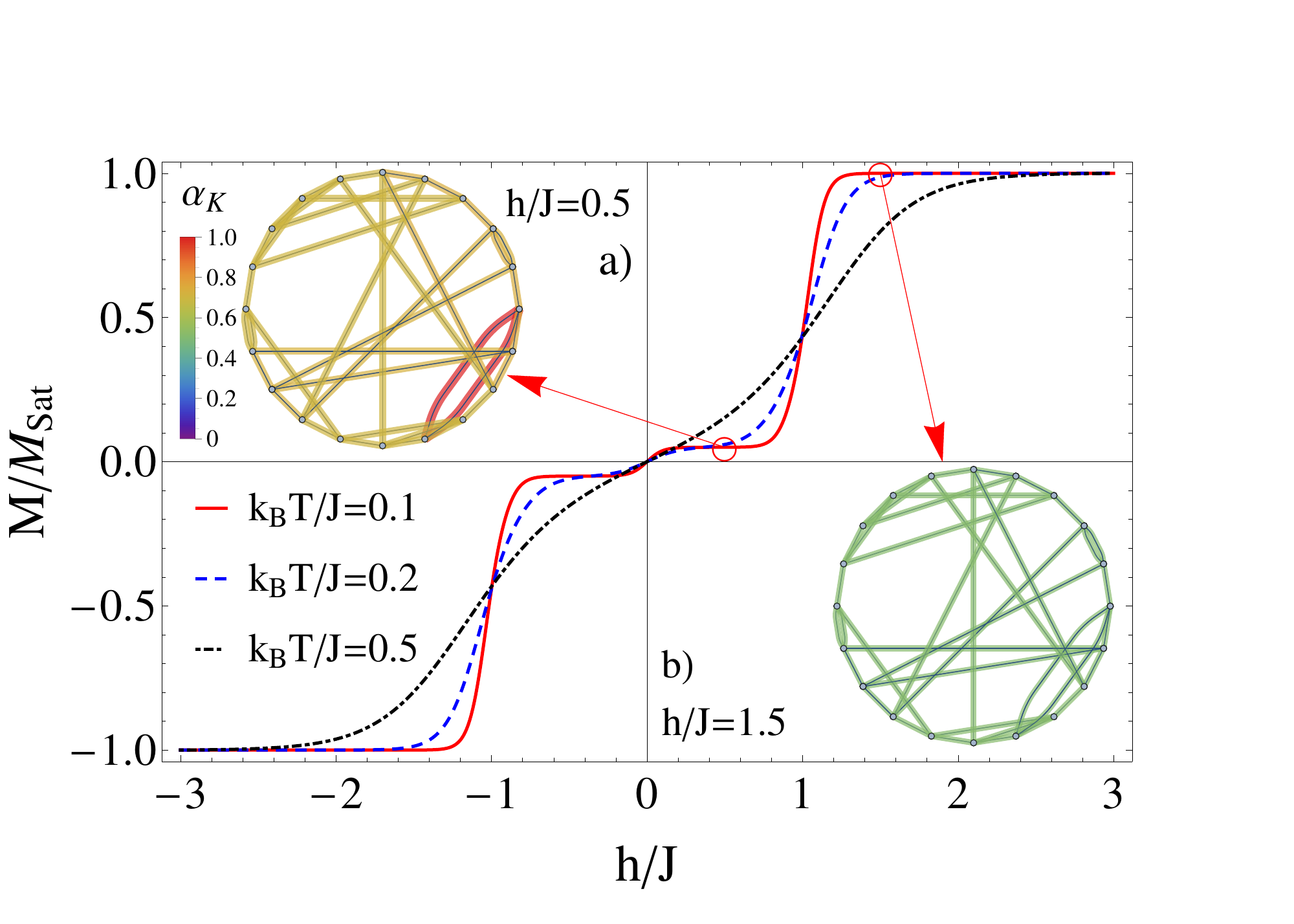}
\caption{(color online). $m_z$ vs. $h/J$ loops of an $N=20$ node 1D
  Ising ring with one randomly distributed long range link per node,
  for $k_BT/J= 0.1, 0.2, 0.5$.  The correlation length $\alpha_K$, as
  a function of cluster size, is given by the colors of the links of
  the a) and b) insets, for $h/J=0.5$ and $1.5$, respectively. }
\label{fig:Magn}
\end{center}
\end{figure}

In order to describe the effects of the topology on the phase
transition we study the correlation length $\alpha_K$ of the $K$-th
closed loop ${\cal C}(n_K)$, with $n_K$ vertices, as 

\begin{equation}
\begin{array}{lll}
F(n_K) & = & \sum_{i,
  \delta\, \subseteq\, {\cal C}(n_K)}
  \left<\sigma_i\sigma_{i+\delta}\right>/n_K \vspace{0.25cm}\\
& = & \sum_{ \delta\,
  \subseteq\, {\cal C}(n_K) } \tr(\sigma M^{\delta}\sigma
  M^{n_K-\delta})/\tr(M^{n_K}) \ ,
\end{array}
\end{equation}
where $F(n_K) $ is the two-point correlation function over a closed
loop ${\cal C}(n_K)$.  On the basis of the above analytic expression,
the correlation length $\alpha_K$ can be obtained from the exponential
dependence \cite{Wipf13} of $F(n_K)\propto e^{ n_K/\alpha_K}$. The
collective behavior of the system is dominated by the local
correlation of magnetically ordered clusters.  In order to show that,
let us consider a 1D Ising ring system with one long range link per
node (over $N=20$ nodes) with a {\it small-world} network structure.
Fig.~(\ref{fig:Magn}) displays the hysteresis loops and cluster
decomposition into local conformations in configuration space.  The
strong correlation of small clusters becomes quite apparent by
inspection of Table~(\ref{table}), which also shows that the magnetic
ordering is robust against thermal fluctuations.

As the magnetic field decreases the correlation length is reduced, and
the magnetic ordering is destroyed by the thermal fluctuations in
large clusters, but it is preserved in small ones. This induces a
residual magnetization, where the continuous line represents
$J_0=-|J|/2\, (J>0)$, and the dashed line the $J_0=|J|/2\,(J<0)$ case.
It is also worth mentioning that this continuous phase transition
resembles ferrimagnetic or canted anti-ferromagnetic ordering, but its
microscopic configuration is created by random long range interactions
through different closed loops, instead of locally induced magnetic
domains.

%
%
\begin{table}
  \caption{Local correlation length $\alpha_K$, for $K  = 2, 7, 8$
      and 23 (where 2+7+8+23 = 40, that is twice the number of nodes)
      as  a function of magnetic field, of a
      1D Ising ring system with  $N=20$ nodes and a {\it small-world}
      network structure. This network is illustrated as the insets of
      Fig.~(\ref{fig:Magn}). The values are normalized to $\alpha_2=1$.}  
\label{table}
\begin{center}
\begin{tabular}{| l | c | c | c | c |}
\hline
\multirow{2}{*}{$h/J$} & \multicolumn{4}{|c|}{Cluster Decomposition} \\
\cline{2-5}
 & $\alpha_2$ & $\alpha_7$ & $\alpha_8$ & $\alpha_{23}$\\
\hline
0.5 &1.0 & 0.72 & 0.74 & 0.70\\
\hline
1.0 & 0.67 & 0.66 & 0.66 & 0.66 \\
\hline
1.5 & 0.51 & 0.51 & 0.51 & 0.51 \\
\hline
\end{tabular}
\end{center}
\end{table}

For low fields the magnetic ordering of large clusters is strongly
suppressed, while small clusters keep their ordering.  This effect is a
consequence of the {\it small-world} structure which breaks up the global
magnetic order, but strongly correlated small clusters survive.

In conclusion, we have shown that random links remove infrared
divergences by breaking the inversion symmetry of a complex network.
This way, the dimensionality restrictions imposed by the MWT, which
assumes translational invariance, are completely removed. Instead, an
unconventional second order phase transition emerges due solely to the
topology of the system. This topologically induced phase transition
displays long range magnetic order in the absence of external fields
through interconnected clusters, as suggested by Gra\ss~et
al.~\cite{Grass15}. Moreover, the {\it small-world} network topology
guarantees that spin correlations are robust against thermal
fluctuations. 


\vskip 8mm
\noindent {\bf Acknowledgments}: This work was supported by the Fondo
Nacional de Investigaciones Cient{\'\i}ficas y Tecnol\'ogicas
(FONDECYT, Chile) under grants \#1150806 (FT), \#1120399 and \#1130272
(MK and JR), \#1150718 (JAV), and CEDENNA through the ``Financiamiento
Basal para Centros Cient{\'\i}ficos y Tecnol\'ogicos de
Excelencia-FB0807''(FT, JR, MK and JAV).

\bibliographystyle{apsrev}

\end{document}